\documentclass[amssymb,amsmath,aps,showpacs,twocolumn,prd]{revtex4}
\usepackage{graphicx}
\def\beq{\begin{equation}}
\def\eeq{\end{equation}}
\def\be{\begin{equation}}
\def\ee{\end{equation}}
\def\bea{\begin{eqnarray}}
\def\eea{\end{eqnarray}}
\def\nnb{\nonumber}

\newcommand{\gsim}{\lower.7ex\hbox{$\;\stackrel{\textstyle>}{\sim}\;$}}
\newcommand{\lsim}{\lower.7ex\hbox{$\;\stackrel{\textstyle<}{\sim}\;$}}

\begin{document}

 \title{ Interaction of cosmic background neutrinos
 with matter of periodic structure }
 \author{ Wei Liao}
 \affiliation{
  Institute of Modern Physics,
 East China University of Science and Technology,\\
 130 Meilong Road, Shanghai 200237, P.R. China
\vskip 0.2cm
  Center for High Energy Physics, Peking University,
 Beijing 100871, P. R. China 
\vskip 0.2cm
Kavli Institute for Theoretical Physics China,\\
Chinese Academy of Sciences, Beijing 100190, P.R. China
}


\begin{abstract}
We study coherent interaction of cosmic background 
neutrinos(CBNs) with matter of periodic structure. 
The mixing and small masses of neutrinos discovered 
in neutrino oscillation experiments indicate that 
CBNs which have very low energy today should be in 
mass states and can transform from one mass state to 
another in interaction with electrons in matter. We 
show that in a coherent scattering process a periodic 
matter structure designed to match the scale of the
mass square difference of neutrinos can enhance the
conversion of CBNs from one mass state to another. 
Energy of CBNs can be released in this scattering 
process and momentum transfer from CBNs to electrons 
in target matter can be obtained. 

\end{abstract}
\pacs{13.15.+g}
 \maketitle

 CBN is one of the major predictions of the big-bang cosmology.
 The big-bang cosmology predicts that the
 temperature of CBNs, if they are relativistic today,
 is around $1.96$ K ~\cite{book}
 ($\sim 10^{-4}$ eV) in the present universe.
 The average number density of CBNs
 per species in the present time is predicted to be~\cite{book}
 \bea
 {\bar n}_{\nu} = 56 ~\textrm{cm}^{-3}. \label{density}
 \eea
 The detection of CBNs is extremely difficult due to the 
 extremely low energy and the small density of CBNs in the present universe.

 In this article we study the scattering 
 of massive CBNs with matter of periodic structure. 
 We point out that a periodic structure of matter, when
 matching with the scale of the mass square difference of massive CBNs, can
 help to convert CBN of one mass state to another and enhance
 the probability of CBNs scattering with matter. 
 Net momentum transfer from
 CBNs to target matter can be achieved in this coherent 
 scattering process. 
 
 In the following we first have a brief review of present 
 knowledge of neutrino masses and mixing.
 Then we study the scattering and conversion of CBNs
 in a target matter of periodic profile.
 We discuss momentum transfer from the wind of CBNs
 to target matter. Finally we conclude.
 
 Neutrino oscillation experiments have shown that neutrinos
 have very small masses and flavor mixing. 
 Two mass square differences of neutrinos have been measured
 in oscillation experiments~\cite{review}:
 \bea
&& \Delta m^2_{21} = 7.50^{+0.19}_{-0.20} \times 10^{-5} ~\textrm{eV}^2, \\
&& |\Delta m^2_{32}| = 2.32^{+0.12}_{-0.08} \times 10^{-3} ~\textrm{eV}^2.
 \label{mass2}
 \eea
 Flavor mixing of neutrinos is expressed as
 \bea
 \nu_l =\sum_i ~U_{li} ~\nu_i,
 \eea
 where $\nu_l$ is neutrino state in flavor base and $\nu_i$
 neutrino state in mass base. Neutrino mixing matrix $U$ can be
 parameterized using three mixing angles $\theta_{12,23,13}$
 and a CP violating phase~\cite{review}.
 In solar, atmospheric and long baseline neutrino 
 oscillation experiments two mixing angles have been measured:
 $\sin^2 2 \theta_{12}\approx 0.86$, ~ 
 $\sin^2\theta_{23} \approx 0.50$~\cite{review} .
 Recent observation of oscillation of reactor anti-neutrinos
 in Daya Bay experiment~\cite{theta13}, confirmed by RENO 
 experiment~\cite{theta13-2}, shows that 
 $\sin^2 2\theta_{13}\approx 0.092$. According to all these
 measurements of neutrino oscillation
 we can conclude that no element of $U$ is zero.
 
 Neutrino masses are also measured in $\beta$ decay experiment and
 in cosmological observations. 
 $\beta$ decay experiment gives an upper bound on the
 mass of electron anti-neutrino~\cite{review}
 \bea
 m_{\bar \nu_e} \lsim 2.0 ~\textrm{eV}
 \label{constraint1}
 \eea
 The observations of the anisotropy of Cosmic Microwave Background
 Radiation and the Sloan Digital Sky Survey give a constraint
 on the total mass of neutrinos~\cite{review}
 \bea
 \sum_i  ~m_i \lsim 0.8 ~\textrm{eV}
 \label{constraint2}
 \eea
 Taken all these measurements of neutrino masses into account, 
 possible patterns of neutrino masses are: Normal Hierarchy(NH)
 with $m_1 \ll m_2 < m_3$, $m_2\approx \sqrt{\Delta m^2_{21}} 
 \approx 0.9 \times 10^{-2}$ eV and 
 $m_3 \approx \sqrt{|\Delta m^2_{32}|}\approx 0.05$ eV; Inverted Hierarchy
 with $m_3 \ll m_1 < m_2$, $m_{1,2} \approx \sqrt{|\Delta m^2_{32}|}
 \approx 0.05$ eV;
 Quasi-Degeneracy(QD) with $m_1\approx m_2 \approx m_3 \lsim 0.3$ eV, 
 $m_i \gg \sqrt{|\Delta m^2_{32}|}$.
 
 Massive CBNs with $m \gg 10^{-4}$ eV
 should be non-relativistic in the present time and
 should be in mass states. Massive CBNs with low velocity should
 also be clustered in galactic halos or in cluster halos.
 For different neutrino mass patterns massive neutrinos
 are different and the contents of non-relativistic CBNs
 are different too. According to the above discussion of
 neutrino mass patterns we
 can figure out that the CBNs which are non-relativistic
 today include $\nu_{2,3}$ for NH, $\nu_{1,2}$ for IH
  and all $\nu_{1,2,3}$ for QD. In the following we will
 consider interaction of massive CBNs with matter.

 The interaction of neutrinos relevant to our analysis is the
 interaction which is non-universal in neutrino flavors. 
 Neglecting radiative corrections
 neutral current interaction of neutrinos with matter is
 universal in flavors and is irrelevant 
 to later analysis. 
 The relevant interaction 
 is given by the charged current interaction
 of neutrino with electron:
 \bea
 \Delta {\cal L} &&=-\frac{4G_F}{\sqrt{2}}{\bar \nu}_e\gamma^\mu \nu_e
 {\bar e}_L \gamma_\mu e_L, \label{Lag}
 \eea
 where $e_L$ and $\nu_e$ are the fields of electron and electron 
 neutrino with left-chirality. $G_F$ is the Fermi constant.
 Using (\ref{Lag}) one can find that
 in an un-polarized target of matter at rest the coherent 
 interaction of neutrino with matter gives a potential term to
 electron neutrino:
 \bea
 \Delta {\cal L} &&= -\sqrt{2} G_F N_e {\bar \nu}_e \gamma^0 \nu_e,
 \label{Lag2}
 \eea
 where $N_e$ is the number density of electron in matter.
 (\ref{Lag2}) describes coherent scattering of
 neutrino with matter in which neutrino coherently
 scatters with many electrons in matter. Momentum transfer from
 neutrino, if not zero, is distributed to very large numbers of
 electrons participating actively in the scattering process 
 and momentum transfer to a single electron
 can be taken zero.
 This case is exactly what we study in later discussion. 

 In the mass base (\ref{Lag2}) can be rewritten as
 \bea
 \Delta {\cal L} &&= - V_e ~U_{ej}^* U_{ei} ~{\bar \nu}_j \gamma^0 \nu_i,
 \label{Lag3}
 \eea
 where $i,j=1,2,3$ and $V_e=\sqrt{2} G_F N_e$.
 According to (\ref{Lag3}), neutrino in one mass state $\nu_i$
 can transform to another mass state $\nu_j$ in interaction with
 electrons in matter.
 Consider such a transition $\nu_i \to \nu_j$ in a target of matter. 
 For a uniform incident flux of neutrino $\nu_i$
 the cross section for the $\nu_i-\nu_j$(${\bar \nu}_i-{\bar \nu}_j$) 
 conversion in matter is
 \bea
 \sigma &&=\frac{1}{2 E_i v_i} \int \frac{d^3 k_j}{(2\pi)^3}
 \frac{1}{2 E_j} 2 \pi \delta(E_i-E_j) ~|M|^2 \nnb \\
 && \times \bigg| \int_\Omega d^3x ~V_e(x) U_{ei}^* U_{ej} 
 ~e^{-i({\vec k}_i-{\vec k}_j)\cdot {\vec x}}
 \bigg|^2 , \label{X-section}
 \eea
 where $E_i$ and $E_j$ are the energies of the initial $\nu_i$ and final 
 $\nu_j$ respectively, $v_i$ the velocity of the initial $\nu_i$
 relative to the target, ${\vec k}_i$ and ${\vec k}_j$ the initial
 and final momenta. $\Omega$ is the volume of the target. 
 $|M|^2$ is the matrix 
 element squared: $|M|^2= k^0_i k_j^0 +{\vec k}_i \cdot {\vec k}_j$ for
 unpolarized Dirac type neutrinos.
 In later discussion we will concentrate on
 massive $\nu_i$ CBNs which are non-relativistic and are clustered in 
 present time. Since the direction of motion is changed in clustering
 process but the spin of neutrino is not, the clustered CBNs can be
 considered mixed with left and right helicities.
 So we can take $|{\vec k}_i | \ll k_i^0$ and
 use $|M|^2=k^0_i k^0_j$ in this case. 
 For Majorana neutrino the neutrino and anti-neutrino are identical
 and $|M|^2$ is replaced by $|M|^2=2(k_i^0 k_j^0+{\vec k}_i\cdot {\vec k}_j
 -m_i m_j)$. It is velocity suppressed if $\nu_i$ and $\nu_j$ are
 both non-relativistic. For relativistic $\nu_j$ and non-relativistic $\nu_i$
 one can use the approximation $|M|^2 =2 E_i E_j$.
 For simplicity we will concentrate on Dirac type neutrino
 and use $|M|^2 =E_i E_j$ in later discussion.

\begin{figure}[tb]
\begin{center}
\begin{tabular}{cc}
\includegraphics[scale=1,width=8cm]{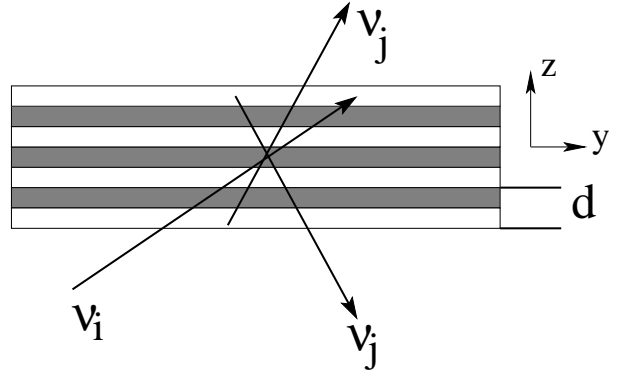}
\end{tabular}
\end{center}
\caption{Scattering of neutrino with matter of periodic structure.}
 \label{reflection}
\end{figure}

 We consider a target of matter which is constant in x and y directions
 and periodic in z direction, as shown in Fig. \ref{reflection}. 
 The potential term in such kind of matter profile satisfies
 \bea
 V_e(z+d) = V_e(z), ~~\frac {d V_e}{dx}= \frac{d V_e}{dy}=0,
 \eea
 where $d$ is the period of the matter profile. 
 A general potential term of 
 such kind periodic structure can be
 expressed using Fourier transformation as
 \bea
 V_e({\vec x})= \sum_n V_n ~e^{i {\vec q}_n \cdot {\vec x}}, \label{profile}
 \eea 
 where $n$ is an integer, ${\vec q}_n= q_n {\hat z}$ and
 $q_n=2 n\pi /d$. $V_n$ satisfies: $V_n^* =V_{-n}$.

 If the path length of neutrino in target is constant the cross-section 
 can be expressed as $\sigma=S p$ where $p$ is the probability 
 of $\nu_i-\nu_j$ conversion and $S$ is the geometric cross 
 section of the target of matter.
 Implementing (\ref{profile}) into (\ref{X-section})
 and integrating over space coordinates
 we find that the conversion probability is
 \bea
 p_n=\frac{|k^z_j||M|^2}{4 E_i^2 v_i E_j} 
 | V_n L_z U_{ej}^* U_{ei}|^2  \frac{4 \sin^2(\Delta_n L_z)}{(\Delta_n L_z)^2},
 \label{prob}
 \eea
 where 
 \bea
 \Delta_n=k_i^z-k_j^z-q_n,
 \eea
 and $L_z$ is the length of target in z direction.
 For ${\bar \nu}_i-{\bar \nu}_j$ conversion the 
 probability is the same.
 We can see that the
 probability is proportional to $|V_n/\Delta_n|^2$ if $|\Delta_n L_z| > 1$.
 If $|\Delta_n L_z| \ll 1$
 the conversion is resonantly enhanced and the probability 
 is proportional to $|V_n L_z|^2$.
 At the resonant point of $\nu_i-\nu_j$ conversion we can find
  \bea
 k_f^x=k_i^x, ~k_i^y=k_j^y, ~k^z_i-k^z_j-q_n=0
 \label{cond2}
 \eea
 This condition of resonant conversion is expressed in short as
 \bea
 {\vec k}_i-{\vec k}_j-{\vec q}_n = 0 \label{cond1}.
 \eea
 When the resonant conversion happens the cross-section
 and the probability $p_n$ are
  proportional to $|V_n|^2$ of one particular $n$ and
 other $V_{n'\neq n}$ effectively contribute zero. 
 
 Using energy conservation we can find that
 \bea
 |{\vec k}_j|=\sqrt{m_i^2-m_j^2+ {\vec k}_i^2} \label{energy1}
 \eea 
 Using $k_i^x=k_j^x$, $k_i^y=k_j^y$ and
 (\ref{energy1}) we can find that
 \bea
 |k^z_j|=\sqrt{m_i^2-m_j^2+(k^z_i)^2}
 \label{energy2}
 \eea

 We will concentrate on massive CBNs which are
 non-relativistic in the present universe. 
 The velocity of massive CBNs at the position of solar
 system depends on the clustering properties of CBNs.
 If CBNs are in virial equilibrium in the local galactic
 halo the velocity of these CBNs is $\sim 200$ km$/$s $\sim 10^{-3} c$
 where $c$ is the speed of light. If CBNs are
 in virial equilibrium in the local cluster halo 
 the velocity of CBNs is a bit larger and
 can reach $\sim 1000$ km$/$s $\sim 10^{-2} c$.
 According to (\ref{constraint2}) we can get $m_i \lsim 0.2-0.3$ eV.
 So we can conclude that $|{\vec k}_i|=|m_i {\vec v}_i|
 \lsim 10^{-3}$ eV and ${\vec k}_i^2 \ll |\Delta m^2_{21,32}|$.
 So using (\ref{energy2}) we get $|k^z_j| \approx \sqrt{\Delta m_{ij}^2}$.
 $|k^z_j|/E_j$ in (\ref{prob}) can be approximated 
 as $|k^z_j|/E_j \approx \sqrt{1-m_j^2/m_i^2}$. Apparently 
 the $\nu_i-\nu_j$ conversion can not happen if $\Delta m_{ij}^2 < 0$.
 For $m_i^2-m_j^2 > 0$, $\nu_i-\nu_j$ conversion is allowed
 and it means that 
 non-relativistic $\nu_i$ converts to $\nu_j$ and releases
 part of its rest energy to kinetic energy of $\nu_j$.

 The transferred momentum
 from neutrino to the target is ${\vec k}_i-{\vec k}_j$.
 Using (\ref{cond1}) we can find that for resonant 
 $\nu_i-\nu_j$ conversion the momentum transfer is ${\vec q}_n$.
 Using (\ref{energy2}) we can find that  $|q_n|= |k^z_j-k_i^z| \approx
 \sqrt{\Delta m^2_{ij}} $ and 
 ${\vec q}_n \approx \pm \sqrt{\Delta m^2_{ij}} {\hat z}$.
 $q_n$ can be positive or negative which correspond to cases that
 $\nu_j$ is reflected or is refracted by the target matter.
 The net momentum transfer per unit time from CBNs is
 \bea
 P= |q_{\pm 1}|  S ~n_i ~\int dv_i ~(p_{+ 1}-p_{-1}) v_i f(v_i), 
 \eea
 where $f$ is the velocity distribution of local $\nu_i$ CBNs, $n_i$ the
 number density of local $\nu_i$ and ${\bar \nu}_i$ CBNs, $S$ the geometric
 cross section of target matter.
 $P$ can be positive or negative which correspond to cases that
 the momentum transfer to target is of positive or negative z direction.
 Apparently there will be no net momentum transfer from CBNs
 to target detector if the probabilities of CBNs being refracted or 
 being reflected by target matter are equal. 
 A periodic structure of target matter
 can make these two probabilities
 differ significantly and net momentum transfer from
 CBNs to target can be obtained.
 
 To illustrate that net momentum transfer can be achieved
 we consider, as an example, the case that $q_{+1,-1}$
 dominate the $\nu_i-\nu_j$ conversion. In such a case 
 the period of the target detector should be arranged to
 satisfy $\frac{2 \pi}{d} \approx \sqrt{\Delta m^2_{ij}}$.
 When neutrino is refracted by the detector 
 $k_j^z=\sqrt{\Delta m^2_{ij} +(k_i^z)^2}$ and $n=-1$.
 We get
 \bea
 \Delta_{-1} =k_i^z+\frac{2\pi}{d}-\sqrt{\Delta m^2_{ij}+(k_i^z)^2}.
 \eea 
 When neutrino is reflected by the detector
 $k_j^z=-\sqrt{\Delta m^2_{ij} +(k_i^z)^2}$ and $n=+1$. We get
 \bea
\Delta_{+1} =k_i^z-\frac{2\pi}{d}+\sqrt{\Delta m^2_{ij}+(k_i^z)^2}.
 \eea 
 We can see that
 $p_{+1}-p_{-1}$ would not be zero and net momentum transfer would be obtained
 if the detector is arranged in such a way that 
 $|\Delta_{+1} L_z|  < 1 < |\Delta_{-1} L_z|$ or
 $|\Delta_{-1} L_z|  < 1 < |\Delta_{+1} L_z|$.
 
\begin{figure}[tb]
\begin{center}
\begin{tabular}{cc}
\includegraphics[scale=1,width=7cm]{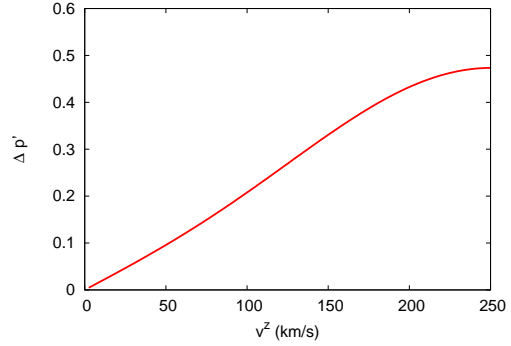}
\end{tabular}
\end{center}
\caption{Asymmetry of $p'_{\pm 1}$ versus $v^z_s$ 
 for $\nu_2-\nu_1$ transition of NH: $m_2=\sqrt{\Delta m^2_{21}}$. 
 $k_s L_z=5$. $k_s=m_2 |{\vec v}_s|$. }
 \label{fig1}
\end{figure}

\begin{figure}[tb]
\begin{center}
\begin{tabular}{cc}
\includegraphics[scale=1,width=7cm]{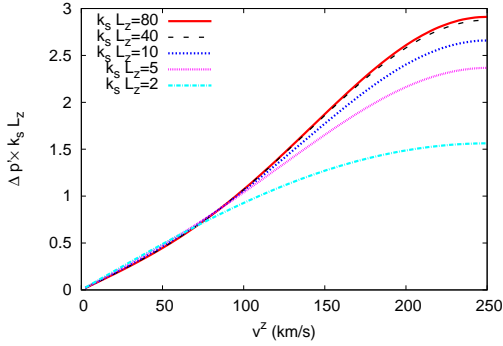}
\end{tabular}
\end{center}
\caption{ $\Delta p' \times k_s L_z$ versus $v^z_s$
 for $\nu_2-\nu_1$ transition and QD: $m_2=\sqrt{|\Delta m^2_{21}|}$.
 Three lines are for $k_s L_z=80, 40$, $10$, $5$ and $2$ separately.
 $k_s=m_2 |{\vec v}_s|$ }
 \label{fig2}
\end{figure}

 To achieve net momentum transfer from neutrino background
 we note that the solar system is moving
 in the local galactic halo or cluster halo. The momentum of CBNs in
 the rest frame of solar system can be written as 
 \bea
 {\vec k}_i= {\vec k}_s+\Delta {\vec k}_i,
 \eea
 where ${\vec k}_s$ is the momentum caused by motion of
 solar system relative to CBN halo. If the structure of
 detector is arranged such that 
 $|q_{\pm1}|= \frac{2\pi}{d}=k^z_s+\sqrt{\Delta m_{ij}^2}+\delta$
 we can find that
 \bea
&& \Delta_{-1}= 2 k_s^z+\Delta k_i^z +\delta 
-\frac{1}{2}\frac{(k_i^z)^2}{\sqrt{\Delta m_{ij}^2}}, \label{delta1}\\
&& \Delta_{+1}= \Delta k_i^z -\delta
+\frac{1}{2}\frac{(k_i^z)^2}{\sqrt{\Delta m_{ij}^2}} \label{delta2},
 \eea
 where $\delta$ is a possible small mismatch between
 $\frac{2 \pi}{d}$ and $k^z_s+\sqrt{\Delta m_{ij}^2}$.
 Since $|{\vec k}_i|/\sqrt{\Delta m_{ij}^2}\lsim 10^{-1}-10^{-2}$ 
 as observed in previous discussions
 the last terms in (\ref{delta1}) and (\ref{delta2}) can be neglected.
 If $ |2 k_s^z L_z| > 1$ it's easy to see that a difference
 between $p_{+1}$ and $p_{-1}$ can be achieved.
 Apparently if $k_s^z=0$ two probabilities should be equal.
 In Fig. \ref{fig1} we can see the asymmetry of $p_{\pm 1}$ clearly
 when $v^z_s$ approaches $250$ km$/$s.
 
\begin{figure}[tb]
\begin{center}
\begin{tabular}{cc}
\includegraphics[scale=1,width=7cm]{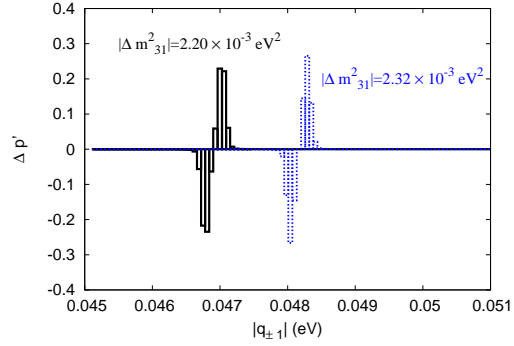}
\end{tabular}
\end{center}
\caption{Asymmetry of $p'_{\pm 1}$ versus $|q_{\pm 1}|$ 
 for $\nu_3-\nu_1$ transition for QD: $m_3=3\times\sqrt{|\Delta m^2_{31}|}$.
 $k_s L_z= 10$. $k_i^z=k_s=m_3 |{\vec v}_s|$. Values of $|q_{\pm 1}|$ for
 resonant conversion varies for different $\Delta m^2_{31}$.
 }
 \label{fig3}
\end{figure}

 In Fig.\ref{fig1} we compute $\Delta p'$:
 \bea
 \Delta p'= \int dv^z_i ~f(v^z_i) ~(p'_{+1}-p'_{-1}),
 \eea 
 where $p_n'=\sin^2(\Delta_n L_z)/(\Delta_n L_z)^2$. The net momentum
 transfer per unit time from CBNs, $P$,
 can be expressed using $\Delta p'$ as
 \bea
 P= |q_{\pm 1}| S n_i c \sqrt{1-{m_j^2 \over m_i^2}}
 |V_{\pm 1} L_z U_{ej}^* U_{ei}|^2 \Delta p',
 \eea
 where $|M|^2\approx E_i E_j$ has been used. 
 As an estimation we use the 
 Maxwellian distribution
 \bea
 f(v)=\frac{1}{\pi^{1/2} a} e^{-(v^z-v^z_s)^2/a^2}.
 \label{distr}
 \eea
 In galactic halo $v^z_s$ can reach $|{\vec v}_s|=250$ km$/$s
 when $z$ direction follows the direction of motion
 of the solar system in Milky Way.
 $a$ is the velocity dispersion of CBNs which is
 taken as $a= 150$ km$/$s in galactic halo.
 In computing Fig. \ref{fig1} we have used $2\pi/d =m_2|{\vec v}_s|
 + \sqrt{\Delta m^2_{21}}$ and $\delta=0$.
 We can see in Fig. \ref{fig1} that the asymmetry of $p'_{\pm 1}$
 disappears when the $v^z_s$ approaches zero and it's maximal
 when $v^z_s=|{\vec v}_s|$.

 $\Delta p'$ does not reach $1$ partly due to the cancellation 
 of $p'_{\pm 1}$. For $|k_s L_z| < 1$ the cancellation can be
 significant. For $|k_s L_z| \sim 1$ we can have $\Delta p'$ 
 of order one. The other factor which reduces $\Delta p'$ is that
 for large $L_z$
 only parts of CBNs with $|\Delta k_i^z L_z| \lsim 1$ contribute to
 the resonant conversion. The larger $L_z$ is, the smaller 
 the fraction of CBNs contributing to resonant conversion.
 $|\Delta p' k_s L_z|$ actually increases as $k_s L_z$ increases and
 it approaches a fixed value for $k_s L_z \gg 1$.
 In Fig. \ref{fig2} we can see this effect.
 We can find that for 
 $k_sL_z=80$ and $40$ the two curves of $\Delta p' k_s L_z$ converge. 
 In this case $\Delta p=p_{+1} - p_{-1}$ 
 approaches to a value which is approximately proportional to $3 L_z/k_s$

 Note that the Earth orbits the Sun with a speed $v_o=30$ km$/$s
 and $v_E$, the speed of the Earth relative to the local halo, varies
 in the range $[v_s-v_o,v_s+v_o]$. As a consequence, $\Delta_{\pm 1}$
 are modulated by ${\vec v}_o$ if considering CBNs interacting
 with a detector on the Earth. In the case $k_s L_z \gg 1$ 
 it's easy to figure out that the range of CBN distribution 
 relevant to resonant conversion, which gives 
 $|\Delta k_i^z L_z| \lsim 1$, is shifted by velocity
 up to $\pm v_o$ 
 in modulation. Using Eq. (\ref{distr}) with $v^z_s$ 
 replaced by $v^z_E$ one can find that for
 $2\pi/d=m_i v_s+\sqrt{\Delta m^2_{ij}}$ the probability
 is changed by a factor in a range $[e^{ -v_o^2/a^2},1]$
 which corresponds to modulation of probability around $4\%$ for
 $a=150$ km$/$s. One can also
 see that if the detector
 is designed such that $2\pi/d=m_i (v_s+v_o)+\sqrt{\Delta m^2_{ij}}$
 the probability is changed by a factor in a range $ [e^{ -(2 v_o)^2/a^2},1]$
 which corresponds to modulation of probability around $15\%$. The
 amplitude of modulation depends on $2\pi/d$ and varies
 from about $4\%$ to about $15\%$.
 Result of numerical computations confirms this estimate. 
 When $k_s L_z \sim 1$ the range of $v^z$
 contributing to resonant conversion is broad
 and a shift in velocity of order $v_o$ does not change much the 
 result. In this case the modulation is weak.

 We note that to achieve maximal $\Delta p'$ sufficiently good
 matching between $|q_{\pm 1}|$ and $\sqrt{\Delta m^2_{ij}}\pm m_i |{\vec v}_s|$
 is needed. Unfortunately we do not have very precise knowledge of
 $\Delta m^2_{ij}$. With (\ref{mass2}) it's hard to achieve a matching 
 with precision to one of a hundred or one of a thousand. To overcome
 this problem one can use a number of detectors which makes a
 scan of the range of $\Delta m^2_{ij}$. For example one can use
 a hundred copies of detectors with identical $L_z$ and slightly different
 $|q_{\pm 1}|$. The values of $|q_{\pm 1}|$ are evenly distributed in
 the uncertain range of $\Delta m^2_{ij}$. In Fig. \ref{fig3} we give an
 example for $\nu_3-\nu_1$ conversion and
 QD: $m_3=3\times \sqrt{|\Delta m^2_{31}|}$. $\nu_3$ is considered heavier
 than $\nu_1$ in this example. The range of $|q_{\pm 1}|$ in this figure
 corresponds to $|\Delta m^2_{31}|$ in the uncertain range 
 $[2.08, 2.68] \times 10^{-3}$ eV$^2$.
 We see that the values of $|q_{\pm 1}|$ which give resonant conversion
 are different for different values of $|\Delta m^2_{31}|$
 and for each $|\Delta m^2_{31}|$ there are $6-8$ values of
 $|q_{\pm 1}|$ of total one hundred which give resonant enhancement.
 Positive or negative $\Delta p'$ correspond to the cases that
 the momentum transfer to target is of the positive or negative
 z direction. We note that by carefully adjusting the
 period of target matter the momentum transfer from CBNs can
 be of the positive or negative direction of the CBN wind. 
 
\begin{figure}[tb]
\begin{center}
\begin{tabular}{cc}
\includegraphics[scale=1,width=7cm]{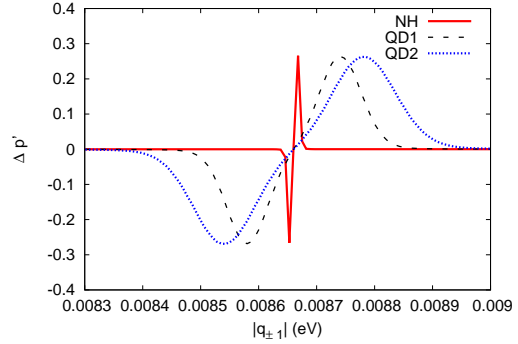}
\end{tabular}
\end{center}
\caption{Asymmetry of $p'_{\pm 1}$ versus $|q_{\pm 1}|$ 
 for $\nu_2-\nu_1$ transition. $k_s L_z=10$. $k_i^z=k_s=m_2 |{\vec v}_s|$.
 NH: $m_2=\sqrt{\Delta m^2_{21}}$; QD1: $m_2=2 \times \sqrt{|\Delta m^2_{31}|}$;
 QD2: $m_2=3\times\sqrt{|\Delta m^2_{31}|}$. }
 \label{fig4}
\end{figure}

 The number of $|q_{\pm 1}|$ which gives resonant conversion
 depends on the neutrino mass pattern. For QD mass pattern the
 neutrino mass is larger than that of NH and IH. The initial
 momentum of CBNs are larger too. So it's easier for QD to achieve
 a matching of $|q_{\pm 1}|$ and $\sqrt{\Delta m^2_{ij}} \pm k_s$
 to the precision of $k_s$. In Fig. \ref{fig4} we can see this clearly in
 $\nu_2-\nu_1$ conversion.
 The range of $|q_{\pm 1}|$ in this figure corresponds to 
 $\Delta m^2_{21}$ in the uncertain range $[6.90, 8.17] \times 10^{-5}$ eV$^2$.
 We see that for QD the resonant region is broad. But for
 NH the resonant conversion happens in a narrow region of $|q_{\pm 1}|$.

 For matter on Earth we find that $V_e \sim 10^{-13}$ eV.
 A detector can be designed to have $V_{\pm 1}$ of the same
 order of magnitude of $V_e$.
 For $L_z= 1$ cm we have $|V_{\pm 1} L_z| \sim 10^{-8}$.
 When sufficiently large $\Delta p'$ is achieved we can have
 a rough estimate of the net momentum transfer
 per unit time from CBNs:
 \bea
 P\sim  && |q_{\pm 1}| \sqrt{1-{m_j^2 \over m_i^2}}
 \frac{S}{1 ~\textrm{m}^2} \frac{n_i}{100 ~\textrm{cm}^{-3}} \nnb \\
 && \times \bigg( \frac{L_z}{1 ~\textrm{cm}} \bigg)^2 \frac{2}{k_s L_z}
 ~\textrm{s}^{-1}.
 \label{momentum}
 \eea 
 In (\ref{momentum}), $\Delta p'\sim 2/(k_s L_z)$ for $k_s L_z >1$
 has been used. This momentum transfer increases as $L_z$ increases
 and as can be seen in (\ref{momentum}) it approaches
 to a value proportional to the volume of the target when $k_s L_z >1$.
 The momentum transfer to target matter gives rise to a mechanical
 force exerted on the target detector. 
 The acceleration due to this force is $a_P=\frac{P}{M}$ where $M$
 is the total mass of the target. Taking $M= \rho S L_z$ where
 $\rho$ is the average density of target and using
 (\ref{momentum}) we can find that that $a_P$ approaches to a constant value
 when $L_z > 1/k_s$:
 \bea
 a_P \sim \frac{|q_{\pm 1}|} {\rho \times 10^4 ~\textrm{cm}^3} 
 \sqrt{1-{m_j^2 \over m_i^2}} \frac{n_i}{100 ~\textrm{cm}^{-3}}.
 \eea
 For example, for $\nu_3-\nu_1$ transition of NH we can find that
 $m_3 \approx 0.05$ eV and $k_s \times \textrm{1cm} \approx 2$. 
 In this case $|q_{\pm 1}|\approx 0.05$ eV$/$c
 and we find that the acceleration 
 can reach $10^{-28}$ cm s$^{-2}$ for $\rho= \textrm{a few} 
 ~\textrm{g}/\textrm{cm}^3$. For other neutrino mass pattern the estimate
 of the momentum transfer and the acceleration is similar except that 
 for $\nu_2-\nu_1$ transition of NH $L_z \gsim 2$ cm 
 should be taken to make $k_s L_z >1$.  

 As can be seen the net momentum transfer to matter is very small.
 The mechanical force exerted on the target is also very small. 
 The acceleration due to this force is far less than the 
 acceleration that can be detected in modern technology,
 that is about $10^{-12}$ cm s$^{-2}$ ~\cite{tech}.
 Moreover, it might be much smaller than the possible mechanical force 
 exerted by dark matter~\cite{dgn}. Detecting this momentum transfer 
 from CBNs using mechanical force is not possible at the moment.
 We should think about other mechanisms to detect the momentum
 transfer from the wind of CBNs. We note that the mechanism considered
 in this article is only sensitive to the charged interaction of neutrino
 with electrons which is non-universal in neutrino flavors. 
 The momentum is transferred from CBNs to electrons in matter 
 through coherent scattering process.
 Considering electrons in conductor or
 superconductor might lead to a better way to detect momentum transfer
 from the wind of CBNs. The research of this topic
 is out of the scope of the present article.

 We note that the mechanism discussed in this article
 make uses of the fact that massive neutrinos can convert
 from one mass state to another in interaction with electrons
 in matter. Momentum is transferred from CBNs to electrons 
 in coherent scattering process with target matter.
 This is different from previous works~\cite{dgn,sl,sbgzk} 
 which considered the coherent scattering of neutrino with 
 nuclei in matter. Another difference from previous works 
 is that the momentum transfer
 from CBNs to target matter discussed in this article
 can be of positive or negative direction of the CBN wind, depending
 on the period of the target matter. Momentum recoil given
 by CBN wind to target matter, discussed
 in ~\cite{dgn,sl,sbgzk}, is always of the same direction of the
 CBN wind.

 We note that if $n_i$ in galaxy is much larger than the average density
 (\ref{density}) the event rate can be enhanced. However, due to 
 Pauli blocking \cite{tg}
 it's difficult for the density of CBNs in our galaxy 
 to be much larger than the value in (\ref{density}) unless the
 neutrino mass reach $\sim 0.5$ eV~\cite{weiler}. Numerical
 simulation of clustering of CBNs does not support $n_i$ in galaxy
 much larger than the average value either~\cite{rw}.
 If considering CBNs clustered in local cluster halo the
 number density is allowed to be much larger and the signal
 of CBNs is larger.
 The mechanism considered in this article can also be
 applied to eV scale sterile neutrino or keV scale sterile
 neutrino dark matter when the period of the target
 matter is designed to be the scale of $\mu$m or nm.

 In conclusion we have studied the coherent scattering of
 CBNs with a detector of periodic matter structure. Massive
 CBNs which are non-relativistic today can convert from
 one mass state to another mass state in interaction with electrons
 in matter.
 Energy of neutrino is released in this scattering process
 and momentum can be transferred from CBNs to target matter.
 We show that a periodic structure of matter can enhance
 the scattering probability when
 the period is matched to the scale of the mass square difference
 of neutrinos. A good arrangement of the
 periodic structure can also select the CBNs to be reflected
 or be refracted by the target matter and lead to net momentum
 transfer to the target matter from the wind of CBNs.
 If a smart way to detect
 this small momentum transfer can be found the result found in this
 article might be useful for designing a realistic detector
 for detecting CBNs in laboratory.

\vskip 0.3cm
 {\bf Acknowledgment:} 
 This work is supported by National Science Foundation of 
 China(NSFC), grant No. 10975052 and grant No.11135009.

\end{document}